# Elon Musk's Twitter Takeover:
# Politician Accounts

*Veli Safak and Aniish Sridhar*

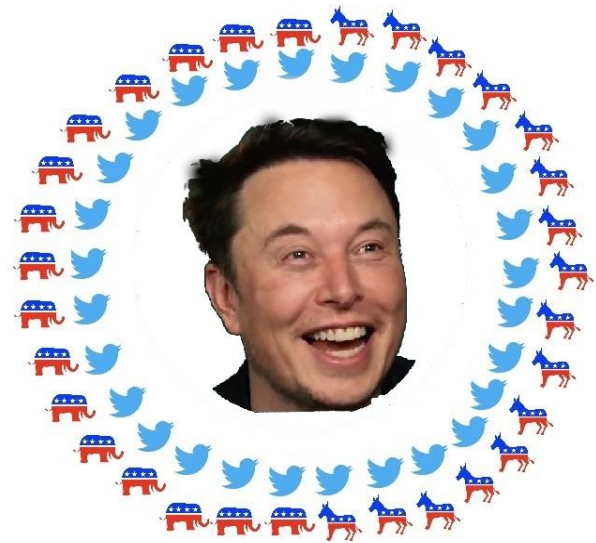

## There and Back Again

On October 26, Elon Musk visited Twitter's headquarter with a sink after updating his Twitter bio to "Chief Twit." A rough road would be an understatement when describing the acquisition process.

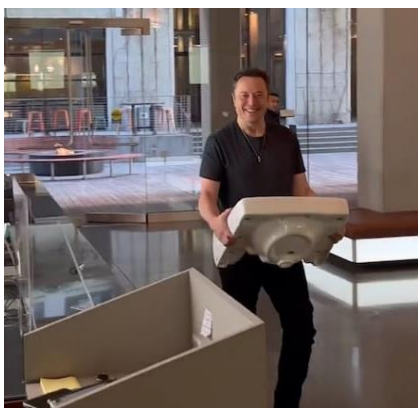

On April 25, Twitter first consented to the takeover. Elon Musk then retracted his offer, citing Twitter's bot issue. Twitter then filed a lawsuit against Elon Musk to enforce the agreement based on his first offer. Elon Musk later changed his mind and completed the deal on October 27.

## Beating Heart of US Politics

According to the most recent data, there are close to 70 million Twitter users in the US. Twitter's recommendation algorithm may be utilized as a political weapon to raise the exposure of particular information. Additionally, it may be used to stop it spreading to millions of users. Numerous studies suggested Twitter's significance in the US politics (see Further Reading).

Democrats vehemently voiced their worries about this acquisition from the beginning to the end.

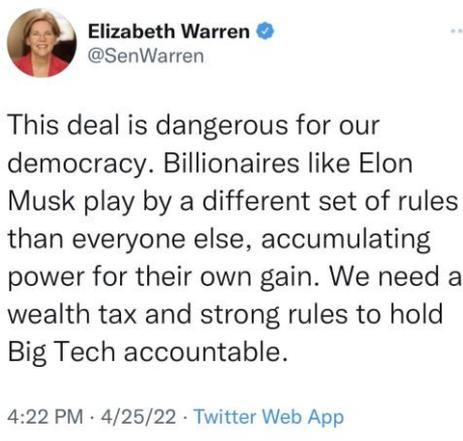
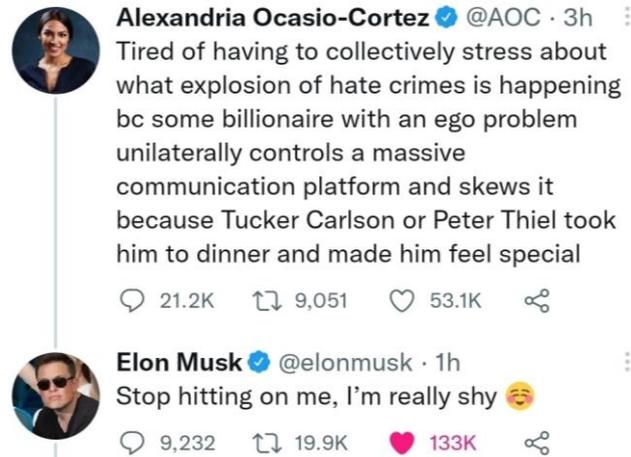

After taking over Twitter, Elon Musk conducted the largest poll on Twitter, asking users if they thought former US president Donald Trump should be allowed back on Twitter. 15,085,458 ballots were cast, with 51.8% of the electorate supporting restoration. Elon Musk decided to restore Donald Trump's account as a result.

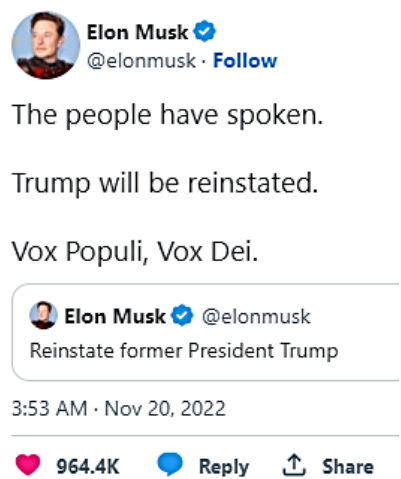

# #ShadowBan

After Twitter accepted Elon Musk's offer, junior US senator from Texas Ted Cruz tweeted about a huge rise in his daily follower numbers. We investigated this assertion and provided quantitative data supporting significant changes between the time Twitter acceptance the offer on April 25 and the time the agreement was finalized on October 27. **Republican politicians saw significant increases** in their follower counts, while **Democrat politicians saw significant decreases**.

## Data

From two major data sources, we created a dataset. First, we utilized the UC San Diego Library's list of official Twitter handles for US Senate and Congress members. Additionally, we gathered follower count data for every available Twitter handle using socialblade.com.

Based on follower counts of 258 Republican and 268 Democrat accounts over 38 days, we examined the effects of the initial agreement and the deal's closure (two sets of 19 days).

**First Set:** April 17 – May 5        **Second Set:** October 17 – November 4

# The First Red Wave (April 26 – May 5)

Between April 18 and May 5, **Republicans gained 2,212,802 followers**. **Democrats lost 31,081 followers** within the same time period. The 26th of April marked the start of a structural change, as the accompanying figure illustrates. **Republicans averaged 18,137 daily followers gain** between April 18 and April 25, while **Democrats averaged 6,691 daily followers**. Between April 26 and May 5, **Republicans averaged 206,771 followers gain** each day while **Democrats lost 8,461 followers** daily.

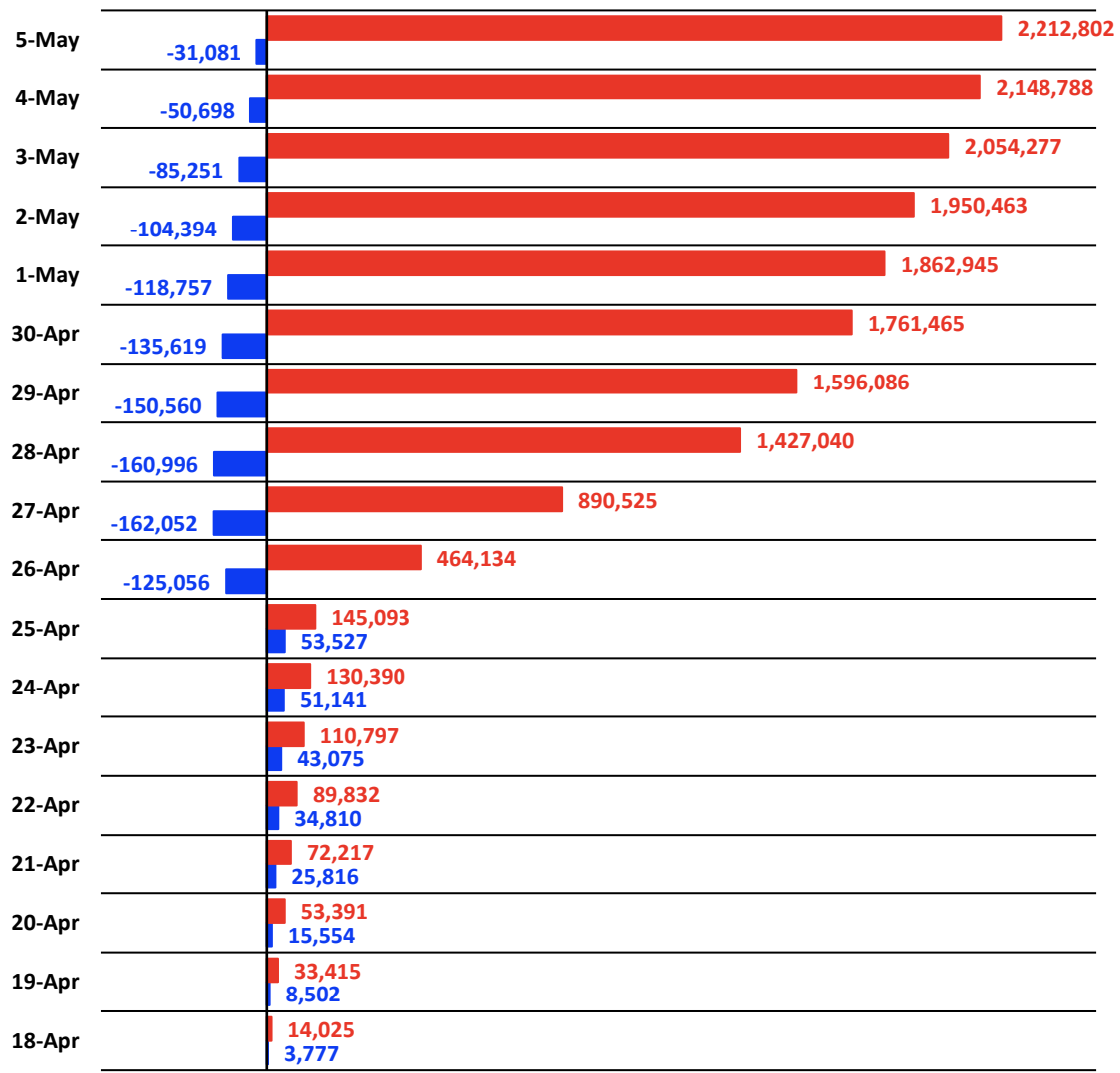

*Figure 1: Cumulative Follower Change Since 17-April*

# The Second Red Wave (October 28 – November 4)

Between October 18 and November 4, **Republicans gained 967,147 followers**. **Democrats lost 544,823 followers** during the same time period. The figure below demonstrates that a structural change started on the 28th of October. **Democrats gained 4,380 followers** per day on average between the October 18 and October 27, while **Republicans gained 14.882 followers** each day. Between October 28 and November 4, **Democrats lost 73,577** while **Republicans gained 102,291 followers** per day on average.

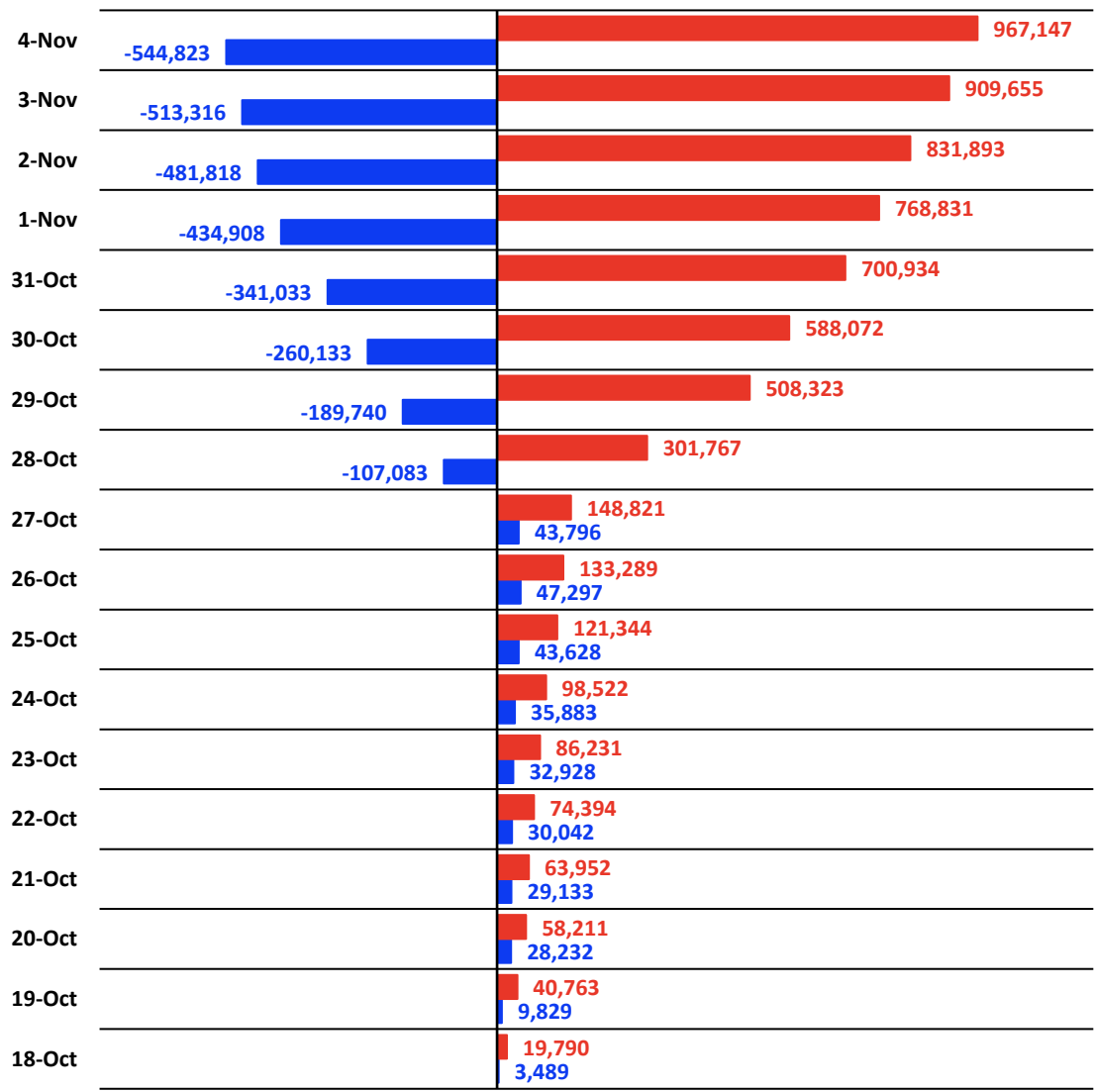

*Figure 2: Cumulative Follower Change Since 17-October*

## Hypothetically Speaking

Following Twitter's acceptance of Elon Musk's offer, we estimate that **an abnormal gain of 1,886,343 followers for Republicans** and **an abnormal loss of 151,517 followers for Democrats** between the 26th of April and the 5th of May. This is based on the average daily change between the 18th and 25th of April.

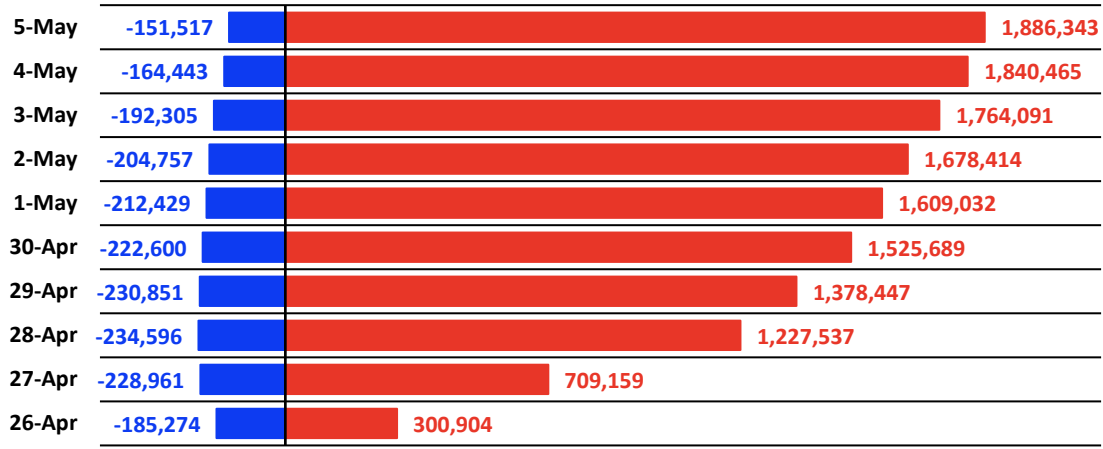

*Figure 3: Cumulative Abnormal Follower Change - First Wave*

To estimate the impact of the acquisition on October 27, we also take into consideration the average daily increase between October 18 and October 27. Estimates indicate that between the 28th of October and the 4th of November after the acquisition, **Republicans saw an abnormal increase of 818,326 followers** while **Democrats experienced an abnormal loss of 588,619 followers.**

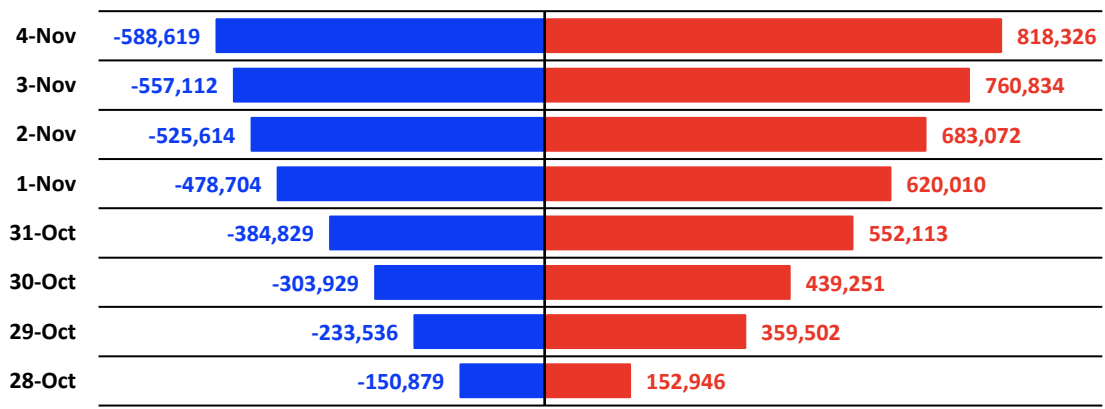

*Figure 4: Cumulative Abnormal Follower Change - Second Wave*

## Conclusion

As of the 4th of November, **Democrats had 55,822,467 followers** and **Republicans had 54,261,683 followers**. **Democrats had 54,232,179 followers** and **Republicans had 48,452,717 followers** on Twitter on April 25, the day Elon Musk's offer was accepted. Although the existence of a shadow ban prior to Elon Musk's acquisition cannot be confirmed merely based on follower count data, the findings indicate that after Twitter accepted Elon Musk's offer, things have changed significantly on the political Twitter. A change in Twitter's recommendation system is one theory for why these changes occurred, but another reason might be a simple rise in Republican supporters' usage following Elon Musk's involvement with the platform.

## About the Authors

Veli Safak is an assistant teaching professor of economics at Carnegie Mellon University Qatar.

Aniish Sridhar is an educational support specialist at Carnegie Mellon University Qatar.

## Further Reading

Buccoliero, L., Bellio, E., Crestini, G. and Arkoudas, A., 2020. Twitter and politics: Evidence from the US presidential elections 2016. *Journal of Marketing Communications*, *26*(1), pp.88-114.

Duncombe, C., 2019. The politics of Twitter: emotions and the power of social media. *International Political Sociology*, *13*(4), pp.409-429.

Howard, P.N., Bolsover, G., Kollanyi, B., Bradshaw, S. and Neudert, L.M., 2017. Junk news and bots during the US election: What were Michigan voters sharing over Twitter. *CompProp, OII, Data Memo*, *1*.

Yaqub, U., Chun, S.A., Atluri, V. and Vaidya, J., 2017. Analysis of political discourse on twitter in the context of the 2016 US presidential elections. *Government Information Quarterly*, *34*(4), pp.613-626.

## Appendix: Infographics

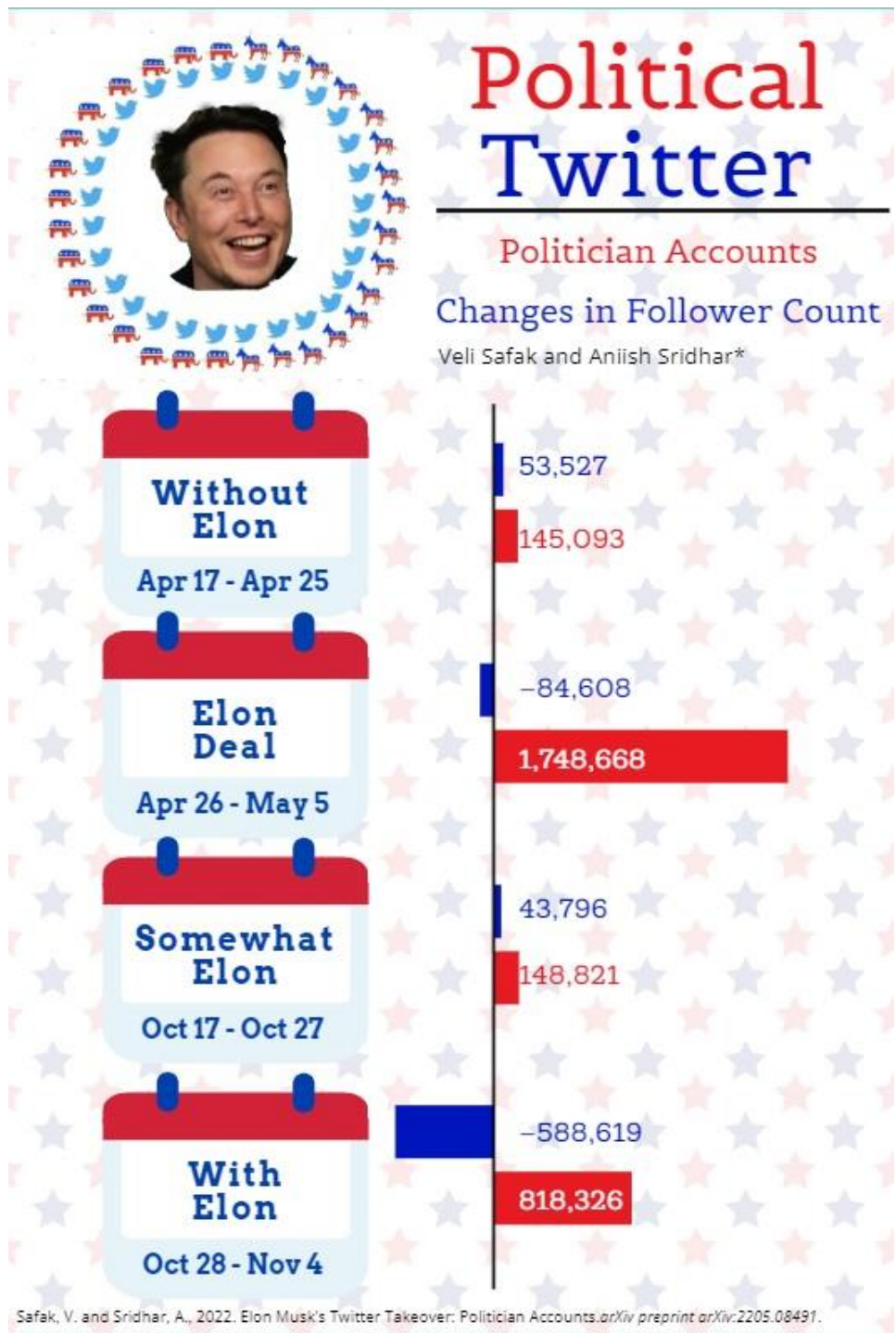